\documentclass[11pt]{article}
\usepackage{a4wide}
\usepackage{amsmath,amssymb,amsfonts}
\usepackage{comment}
\usepackage{epsfig,verbatim,graphics}
\usepackage{subfigure,slashed}

\newcommand{\mathsym}[1]{{}}

\newcommand{\rmi}{\mathrm{i}}
\newcommand{\eref}[1]{(\ref{#1})}
\newcommand{\erefdbl}[2]{(\ref{#1}, \ref{#2})}
\renewcommand\({\left(}
\renewcommand\){\right)}
\renewcommand\[{\left[}
\renewcommand\]{\right]}

\newcommand{\dd}{{\rm d}}
\declareslashed{}{{^-}}{.1}{0.1}{\rm d}

\newcommand{\e}{{\rm e}}

\newcommand\eps{\epsilon}
\newcommand\mpl{M_{\rm P}}
\newcommand\mplvar[1]{M_{\rm P,#1}}

\def\be{\begin{equation}}
\def\ee{\end{equation}}

\def\mcL{\mathcal{L}}
\def\mcO{\mathcal{O}}

\def\nn{\nonumber}
\def\({\left(}
\def\){\right)}

\newcommand{\roughly}[1]{\mathrel{\raise.3ex\hbox{$#1$\kern-0.85em
\lower1ex\hbox{$\sim$}}}}

\begin{document}

\begin{titlepage}
\begin{center}

\today
\hfill NIKHEF 2013-033\\
\hfill DAMTP-2013-61

\vskip 1.5cm

{\LARGE \bf Quantum corrections in Higgs inflation:\\ the real scalar case}

\vskip 1cm

\renewcommand*{\thefootnote}{\fnsymbol{footnote}}
\setcounter{footnote}{0}

{\bf
    Damien P.\ George$^{1,2}$\footnote{{\tt dpg39@cam.ac.uk},
        \footnotemark[2]{\tt sander.mooij@ing.uchile.cl},
        \footnotemark[3]{\tt mpostma@nikhef.nl}},
    Sander Mooij$^{3,4}$\footnotemark[2] and
    Marieke Postma$^3$\footnotemark[3]
}

\renewcommand*{\thefootnote}{\number{footnote}}
\setcounter{footnote}{0}

\vskip 25pt

{\em $^1$ \hskip -.1truecm
Department of Applied Mathematics and Theoretical Physics,\\
Centre for Mathematical Sciences, University of Cambridge,\\
Wilberforce Road, Cambridge CB3 0WA, United Kingdom
}

\vskip 20pt

{\em $^2$ \hskip -.1truecm
Cavendish Laboratory, University of Cambridge,\\
JJ Thomson Avenue, Cambridge CB3 0HE, United Kingdom
}

\vskip 20pt

{\em $^3$ \hskip -.1truecm
Nikhef, \\Science Park 105, \\1098 XG Amsterdam, The Netherlands
}

\vskip 20pt

{\em $^4$ \hskip -.1truecm
FCFM, Universidad de Chile, \\Blanco Encalada 2008, \\Santiago, Chile
}

\end{center}

\vskip 0.5cm

\begin{center} {\bf ABSTRACT}\\[3ex]
\end{center}
We present a critical discussion of quantum corrections,
renormalisation, and the computation of the beta functions and the
effective potential in Higgs inflation.  In contrast with claims in
the literature, we find no evidence for a disagreement between the
Jordan and Einstein frames, even at the quantum level.  For clarity of
discussion we concentrate on the case of a real scalar Higgs.  We
first review the classical calculation and then discuss the back
reaction of gravity.  We compute the beta functions for the Higgs
quartic coupling and non-minimal coupling constant.  Here, the
mid-field regime is non-renormalisable, but we are able to give an
upper bound on the 1-loop corrections to the effective potential.  We
show that, in computing the effective potential, the Jordan and
Einstein frames are compatible if all mass scales are transformed
between the two frames.  As such, it is consistent to take a constant
cutoff in either the Jordan or Einstein frame, and both prescriptions
yield the same result for the effective potential.  Our results are
extended to the case of a complex scalar Higgs.

\end{titlepage}

\newpage
\setcounter{page}{1} \tableofcontents

\newpage


\section{Introduction}

In the paradigm of Higgs inflation the standard model (SM) Higgs
doublet plays the role of the inflaton, and provides an
almost-constant de Sitter vacuum energy to exponentially expand the
universe during its initial stages~\cite{salopek,bezrukov1}.  The
model introduces a single new term to the standard model with gravity,
to couple the Higgs doublet $\Phi$ to the Ricci scalar, $\xi R
\Phi^\dagger \Phi$ \cite{fakir,kaiser94}.  The associated
dimensionless coupling constant $\xi$ must necessarily be large,
$\xi\sim2\times10^4$ to give slow roll inflation in agreement with
data from the cosmic microwave background (CMB).  The classical
analysis of Higgs inflation is well understood: one transforms to the
Einstein frame to make the gravity sector canonical, redefines the
Higgs degree of freedom (in unitary gauge) to obtain a canonical
kinetic term, and uses the resulting potential to compute slow roll
parameters in the usual way.  This gives a connection between the
quartic coupling in the Higgs potential and the new non-minimal
coupling, and parameters of the CMB, $n_s$ and $r$.

Higgs inflation is a theory spanning many orders of magnitude in
energy.  The couplings in the SM are measured by collider experiments
at energies around the electroweak scale, whilst inflation and its
observables are defined at the inflationary scale, 13 orders of
magnitude higher.  This disparity of scales means that the leading
logarithmic corrections due to quantum loops will be large, or,
equivalently, that the running of the couplings is significant from
the electroweak to the inflationary scale.  It is therefore important
to consider quantum effects in Higgs inflation, but in doing so many
problems arise, both conceptual and technical
\cite{bezrukov2,bezrukov_1loop,bezrukov_2loop,
  bezrukov:consistency,wilczek,barvinsky,barvinsky2,barvinsky3,lerner,allison}.

The main quantity of interest is the loop-corrected effective
potential of the Higgs/inflaton.  As previously mentioned, knowledge
of the potential allows one to compute slow roll pa\-ra\-me\-ters, and
incorporating loop-corrections and running of the couplings allows one
to obtain a more accurate answer.  The different degrees of freedom in
the SM all provide competing corrections, along with corrections due
to Hubble expansion; also there is the important fact that the Higgs
quartic coupling runs very close to zero.  The transformation of the
theory from the original Jordan frame where the theory is defined, to
the Einstein frame with canonical gravity, leads to additional issues
depending on which frame the quantisation is performed in.  Quantising
in the Jordan frame with gravity a non-dynamical background leads to
incorrect beta functions for the running couplings, unlike in the
Einstein frame where non-dynamical gravity introduces only a small
error.  Renormalising the theory in either frame with a constant UV
cutoff can lead to different results if not done correctly.  Finally,
the model itself is inherently non-renormalisable, and so at certain
points in field space it seems impossible to compute loop corrections
to the potential.

For Higgs inflation as a theory to make accurate predictions it is
crucial to sort out all of the above-mentioned issues in a consistent
and rigorous way.  It is the aim of this paper to address these issues
and give a conservative set of answers and associated formulae.  We
will discuss the following points.
\begin{itemize}
\item At the classical level, the conformal transformation relating
  the Jordan and Einstein frames is just a redefinition of the fields,
  and the two frames are physically equivalent. The two- and
  three-point functions of the perturbations during inflation are the same
  in both frames \cite{kaiser95,flanagan,janW1,janW2,white1,white2}.  We shall
  demonstrate that this equivalence also holds for the 1-loop effective
  potential, and, derived from that, for the beta functions.  The Jordan or
  Einstein frame (or any other frame) is not fundamentally better, it
  is just easier in some frames to compute certain quantities.

\item In the literature it is argued that calculating the effective
  potential in the Jordan frame gives a field-dependent cutoff in the
  Einstein frame and vice versa
  \cite{bezrukov_1loop,bezrukov_2loop,barvinsky,barvinsky3,allison}.
  Moreover, these two prescriptions are then claimed to lead to two
  different effective potentials.  We show explicitly that in both
  frames the dependence on the cutoff can safely be eliminated and
  that both frames give the same result.  Therefore, the theory does
  not depend on the choice of the cutoff, or on the UV
  completion.\footnote{There is some small sensitivity to the UV
    theory due to non-renormalisability in the mid-field regime, but
    this effect is subdominant when computing the effective potential,
    and we can safely ignore it.}

\item One item that is expressed differently between the two frames is
  the adiabatic scalar degree of freedom. Indeed, it is a different
  mixture of the Higgs field and the scalar degree of freedom in the
  metric the Jordan frame than in the Einstein frame. If one treats
  gravity as a classical background and only quantises the Higgs
  field, this thus amounts to different approximations. The
  approximation is fine for the Einstein frame, where the error made
  is of the order of the slow roll parameter $\eps$ and therefore
  subleading. However, we will claim that in the Jordan frame this
  approximation is not at all valid. As a result, the renormalisation
  group equations (RGEs) thus found in the Jordan frame
  \cite{bezrukov_1loop,tranberg,buchbinder,kirsten,barvinsky,barvinsky3,lerner}
  are incorrect. Note in this respect that also the RGE for the
  non-minimal couplings used in \cite{
  wilczek} treats
  gravity classically in the Jordan frame.

\item The theory is renormalisable in the small field regime (in the
  usual sense of low energy effective field theories) and in the large
  field regime (thanks to an approximate shift symmetry
  \cite{bezrukov:consistency,ferrara2}).  In the mid-field regime the theory is
  most likely non-renormalisable, but we are still able to compute the
  leading order contribution to the RGEs in this range, being as
  conservative as possible.  Thus, the potentially infinite number of
  counter terms that need to be introduced to properly renormalise the
  theory are parametrically suppressed, and we can still make
  predictions.

\item The Higgs doublet has four scalar degrees of freedom, and the
  non-minimal coupling to gravity means that these scalars are
  inherently mixed (their quadratic terms in the action are not
  simultaneously diagonalisable \cite{kaiser}).  We show how to
  include such degrees of freedom in quantum loops by diagonalising
  the mass matrix, for a fixed point in field space, to find the
  physical masses for the 1-loop contribution to the effective
  potential. Both the contribution of the Higgs and the (massive)
  Goldstone bosons are suppressed during inflation.  This is in
  contrast with the claims made in
  Refs.~\cite{barvinsky3,lerner,allison} that only the Higgs field is
  suppressed.

\item 
  Due to the large non-minimal coupling $\xi$, unitarity of
  $\text{SU}(2)$ gauge boson scattering is lost below the inflationary
  scale in the small and mid-field regimes
  \cite{barbon,cliff1,cliff2}. The unitary bound is field dependent,
  and it has been shown that in all three regimes the typical energy
  scales in Higgs inflation are parametrically lower than the scale of
  unitarity violation \cite{bezrukov:consistency,ferrara2}. That may
  already solve the problem of non-unitarity of Higgs inflation, but new physics still seems required to restore unitarity in the SM vacuum. However, we think that the situation is actually even better. We shall briefly
  comment on the possibility that this new physics necessarily appears
  only at Planck-scale energies.  This gives further hope that the
  predictions of Higgs inflation can be trusted.

\end{itemize}

As the issues involved are largely conceptual, we start with the simplest
setting, that of a real scalar field with a non-minimal coupling, and
extend to a complex field in the latter sections of the paper.   Since
there is no gauge sector, the anomalous dimension of the Higgs vanishes
and we also do not have gauge or (fermionic) Yukawa couplings to deal with.
Only the correction to the effective potential needs to be calculated.
The 1-loop effective potential can be expressed in terms of the masses of the
fields, i.e.\ only depends on the action at the quadratic level, and is not
hampered by the complicated non-minimal kinetic terms.  Since the Higgs is
light during inflation, its contribution should be calculated in
FLRW; in other regimes a Minkowski approximation suffices.  Even with
these simplifications it is still possible to discuss and address all
of the above-mentioned issues.  In a follow up paper we aim to build upon
the results obtained here and look at the full standard model.

It is important to note that a Higgs mass $m_h=125~ \text{GeV}$,
compatible with the measurements by ATLAS and CMS \cite{atlas,cms},
leads to the Higgs quartic coupling $\lambda$ running negative at
energies below the inflationary scale
\cite{lambdarun1,lambdarun2}. This is for central values of the top
mass and strong coupling, and can be avoided, albeit marginally, by
taking $m_t\sim171~ \text{GeV}$, $3\sigma$ lower than the central
value \cite{lambdarun3}.  One could also extend the minimal
Higgs inflation model to include additional particles that modify the
running of $\lambda$ \cite{lebedev,miro}.  Whatever the specifics, the
results presented here can be applied to such a model, and can also be
applied more generally to models of inflation where the inflaton
scalar is not the Higgs doublet, but is still non-minimally coupled to
the Ricci scalar. We do assume, though, that the non-minimal coupling
is large $\xi \gg 1$, and can be used as an expansion parameter.

The paper is organised as follows.  We begin in
Sec.~\ref{sec:classical} by reviewing the well established classical
calculation, and discuss the back reaction from gravity.  In
Sec.~\ref{sec:beta} we find the beta functions for the quartic coupling
and non-minimal coupling $\xi$, including a conservative bound on the
running of the couplings in the non-renormalisable mid-field regime.
In Sec.~\ref{sec:jvse} we discuss the equivalence of the Jordan and
Einstein frames when computing the effective potential.  We show how
the Jordan and Einstein frames should give a result independent of
which frame the cutoff is chosen to be constant in.  In addition, we
argue that in the Einstein frame gravity can be treated as a
non-dynamical background and only introduces a small error, in
contrast with the Jordan frame.  Our results are extended to a complex
scalar Higgs in Sec.~\ref{sec:complex}, and we make some brief
remarks in Sec.~\ref{sec:unitarity} on the unitarity bound.
Finally, Sec.~\ref{sec:discussion} compares our results with the
literature and draws conclusions.


\section{Higgs inflation, the classical background}
\label{sec:classical}

Higgs inflation takes the SM Lagrangian with the Einstein-Hilbert term
for gravity and adds a non-minimal coupling between the Higgs doublet
$\Phi$ and the Ricci scalar $R$
\cite{salopek,bezrukov1,fakir,kaiser94}.  The Jordan frame is the
frame in which the theory is defined, the Lagrangian being (with
$+---$ metric signature)
\be
\mcL_J = \sqrt{-g^J}\[ -\frac12 \mpl^2 \(1+ \frac{2\xi \Phi^\dagger
  \Phi}{\mpl^2} \) R[g^J] + \mcL_{\rm SM}\] .
\ee
A label $J$ denotes a Jordan frame quantity, $\mpl$ is the Planck mass
and $\xi$ is the new dimensionless coupling.
For the purposes of this paper we require only the Higgs sector, and so
drop the gauge fields and fermions.  Thus, the only relevant parts of the
SM Lagrangian are the Higgs kinetic term and the Higgs potential
\be
V_J = \lambda(\Phi^\dagger \Phi - v^2/2)^2 .
\ee

The Einstein frame is obtained by redefining the metric degrees of freedom
such that gravity is in its canonical Einstein-Hilbert form.  As a result,
the non-minimal coupling to gravity is removed and appears instead as
non-canonical kinetic structure in the matter sector.  The appropriate
transformation is a conformal Weyl rescaling, made by
$g_{\mu\nu} = \Omega^2 g^J_{\mu\nu}$, with
\be
\Omega^2 = \(1+ \frac{2\xi \Phi^\dagger \Phi}{\mpl^2}\) .
\label{Omega}
\ee
The new Einstein-frame metric is $g_{\mu\nu}$, and the resulting Lagrangian
is
%
%
\be
\mcL_E =\sqrt{-g}\[ -\frac12 \mpl^2 R[g]
+  \frac{1}{\Omega^2} (\partial_\mu \Phi)^\dagger (\partial^\mu \Phi) + \frac{3 \xi^2}{\mpl^2\Omega^4}
 \partial_\mu (\Phi^\dagger \Phi) \partial^\mu (\Phi^\dagger \Phi) - \frac{V_J}{\Omega^4}\] .
\label{LE}
\ee
We define the Einstein frame potential $V = V_J/\Omega^4$.  The Higgs kinetic
term is non-minimal and mixes the 4 degrees of freedom in the doublet.  The
model can be considered a non-linear sigma model with a non-trivial target/field
space.  Let $\chi_i$ run over all real fields in cartesian space, namely
$\sqrt{2}\Phi^\intercal=(\chi_1+\rmi\chi_2,\chi_3+\rmi\chi_4)$; this generalises
easily to other sized multiplets.  Then the field-space metric in component form
is $\gamma_{ij}$ defined by
\be
\frac{\mcL_E}{\sqrt{-g}} \supset
\frac12 \gamma_{ij} \partial \chi_i \partial \chi_j = \frac12 \[\frac{\delta_{ij}}{
  \Omega^2} + \frac{ 6
\xi^2}{\mpl^2\Omega^4} \chi_i \chi_j\] \partial \chi_i \partial \chi_j.
\label{non_can}
\ee
The curvature on field space is non-zero, $R[\gamma_{ij}] \neq 0$, and
hence there is no field redefinition among the set $\{\chi_i\}$ which
diagonalises the kinetic terms \cite{kaiser}.  This problem has a
partial solution if one expands the fields around a fixed background,
$\chi_i=\chi^0_i+\delta\chi_i$, and diagonalises the kinetic term for
$\delta\chi_i$ at a specific point in field space defined by
$\{\chi^0_i\}$.  This is akin to the situation in general relativity
where, if the space-time curvature is non-zero, one can go locally to
a Minkowski frame, but not globally.  Considering $\Phi$ a real scalar
field with only 1 component this problem disappears, as the kinetic
terms for a single degree of freedom can always be diagonalised.

For the classical analysis of Higgs inflation one considers only the background
radial mode of $\Phi$, call it $\phi$, defined by $\phi^2=2|\Phi|^2$.
This field can be canonically normalised via
\be 
\frac12 \gamma(\phi) (\partial \phi)^2
= \frac{1}{2\Omega^2} \(1+\frac{6  \xi^2}{\mpl^2\Omega^2} \phi^2\) (\partial \phi)^2
= \frac12 (\partial h)^2 .
\ee
The differential equation can be solved for $h$, but the solution is
complicated and does not give much insight (it also cannot be inverted
to get $\phi(h)$).  Instead, we solve it in the three regimes ---
small field, mid field and large field --- and then patch them
together \cite{bezrukov:consistency,ferrara2}.  The value of $\xi$ is
large (as will be shown shortly) and its inverse is used to define the
boundaries of the three regimes.  From now we on we set $\mpl=1$.
\begin{enumerate}
\item Small field regime: $\phi< 1/\xi$.  Then $h'(\phi) \sim 1$, and with
the boundary condition $h(0)=0$ the solution is  
\be
h =\phi.
\ee
The effect of the non-minimal coupling is negligibly small.

\item Mid field regime: $1/\xi< \phi <1/\sqrt{\xi}$.  Then
$h'(\phi) \sim \sqrt{6} \xi \phi$.  
Matching to the small field regime gives the boundary condition
$h(1/\xi) = 1/\xi$, and solution
\be
h = \sqrt{3/2} \xi \phi^2 + (1-\sqrt{3/2})/\xi.
\label{h_mid}
\ee

\item
Large field regime: $\phi>1/\sqrt{\xi}$. Then $h'(\phi) \sim \sqrt{6}/\phi$.
The boundary condition is $h(1/\sqrt{\xi}) = \sqrt{3/2}$, yielding the
solution
\be
h = \sqrt{6} \ln(\phi \sqrt{\xi}) + \sqrt{3/2}
\qquad \Rightarrow \qquad
\phi =\frac1{\sqrt{\xi}} \e^{(h - \sqrt{3/2})/\sqrt{6}} .
\label{h_large}
\ee
\end{enumerate}

In the large field regime the potential can be expanded in 
\be
\delta \equiv \frac{1}{(\xi \phi^2)} \ll 1.
\label{delta}
\ee
 Note that $\delta$ is independent of both $\xi$ and
$\lambda$ when expressed in terms of the canonically
normalised field $h$.  Furthermore, neglecting the small mass term
$v$, the potential only depends on the combination $\bar \lambda
\equiv \lambda/\xi^2$.  We expand all quantities to first order in
$\delta$.  This gives for the kinetic terms $\sqrt{\gamma} =
\sqrt{6}/\phi (1-\delta) + \mcO(\delta^2)$.  Here terms $1/\xi \ll 1$
are neglected, thus the result is only valid for large non-minimal
coupling.  The exponential solution for $h$ given in \eref{h_large} is
only valid to lowest order.  To find the derivatives of the potential
we use $\sqrt{\gamma} V_h = V_\phi$, and similar for higher
derivatives; this takes first order corrections into account.  Up to
$\mcO(\delta^2)$ we find
\be
V= \frac{\bar \lambda}{4}\(1 -2\delta\),\quad
V_h = \frac{\bar \lambda \delta}{\sqrt{6}} (1-2\delta),\quad
V_{hh} = \frac{-\bar \lambda \delta}{3} (1-3\delta),\quad
V_{hhh} = \frac{2\bar \lambda \delta}{3\sqrt{6}} (1-5\delta).
\label{large_Vh}
\ee
Using the standard expressions for the slow roll parameters,
$\eps=(V_h/V)^2/2$ and $\eta=V_{hh}/V$, we obtain
\be
\eps =\frac43\delta^2, \quad
\eta = -\frac43 (\delta - \delta^2) .
\label{large_sr}
\ee
Inflation ends for $\eps \approx 1$ which gives $\delta_{\rm end} \sim
\sqrt{3/4}$.  The number of e-folds is
\be
N  = \int \dd h (2\eps)^{-1/2} = \int \dd \phi \frac{V}{V_\phi} 
\(\frac{\partial h}{\partial \phi} \)^2  \approx \frac34 \xi (\phi_*^2
-\phi_{\rm end}^2) \approx \frac3{4 \delta_*} .
\ee
Setting $N \approx 60$, we find $\delta_* \sim 3/4N \approx 1/80$ when
observable scales leave the horizon.  
The COBE normalisation gives
\be
\(\frac{V}{\eps}\)_* = (0.027)^4 
\qquad \Rightarrow \qquad
\bar \lambda = 4 \times 10^{-10} 
\qquad \Rightarrow \qquad
\xi/ \sqrt{\lambda} = 5 \times 10^4 .
\label{xival}
\ee
Since $\lambda\sim0.1$ in the SM we require $\xi$ of order
$2\times10^4$ to agree with the data.\footnote{Note that if $\lambda$ runs to
a value very close to zero then $\xi$ can be smaller, as discussed in
Ref.~\cite{allison}}
The predictions for the spectral index and tensor-to-scalar ratio are
%
\be
n_s = 1+ 2\eta - 6 \eps \approx 1 - \frac83 \delta_* - \frac{16}3 \delta_*^2
\approx 0.966, \qquad
r = 16\eps = 8 \delta_*^2 = 0.0032 .
\label{nsval}
\ee
These are within current $1\sigma$ bounds from Planck~\cite{planck}.


\subsection{Back reaction of gravity}
\label{sec:backr}

The results used above for the perturbation spectrum take into account
the back reaction of gravity, that is the mixing between the Higgs
field and scalar degree of freedom in the metric. Only during
inflation these gravity corrections are important, i.e.\ in the large
field regime. In this subsection we quickly review the approximation
that takes gravity as a classical background. As we will discuss
later, for the quantum corrections calculated in the Einstein frame
this is a good lowest order approximation, because of a hierarchy in
the slow roll parameters $\eps \ll \eta$; however, for the calculation
in the Jordan frame, it fails.

Consider a canonical real scalar during inflation, and expand it
around a classical background $\sqrt{2}\Phi(\vec x,t) = \phi(t) +
\delta \phi(\vec x,t)$.  We work in conformal coordinates ($\tau$ is
conformal time) using an FLRW metric with scale factor $a$, and rescale
the fields $\hat \phi = a \phi$ etc.
Neglecting the back reaction the quadratic action for the scalar
perturbations is
\be
S^{(\text{quad})} = -\frac12 \int \dd \tau \dd^3 x \delta  \hat \phi
\( \partial^2 + \hat m_{\delta \phi}^2 \) \delta \hat \phi ,
\label{S_ho}
\ee
with effective ``conformal'' mass $ \hat m^2_{\delta \phi}= a^2 m_{\delta \phi}^2$ where
\be
 m_{\delta \phi}^2=   \[V_{\phi\phi} - (\dot H +2 H^2)\] = -
H^2\(2- 3\eta - \eps_H\) .
\label{m1}
\ee
In this section we use the slow roll approximation which implies $\epsilon \approx \epsilon_H$ (with $\epsilon \equiv 1/2 (V_h /V)^2$ and $\epsilon_H\equiv -\dot{H}/H^2$) and $\eta \approx \eta_H$ (with $\eta\equiv V_{hh}/V$ and $\eta_H \equiv \ddot \phi/(H\dot \phi) + \eps_H$). Furthermore we use the third slow roll parameter $\qquad \xi_{(2)} = -\dddot \phi/(H^2 \dot \phi)$.
 One can now quantise the above
action for the harmonic oscillator with a time-dependent mass, and
derive the perturbation spectrum.

We now contrast this with the action when the back reaction of gravity
is included.  The effective scalar degree of freedom during inflation
is a mixture of a metric degree of freedom and the Higgs field, which
can be expressed in gauge invariant form via the Sasaki-Mukhanov
variable \cite{mukhanov,mukhanov2,sasaki}
\be
v = a \(\delta \phi + \frac{\dot \phi}{H} \psi\).
\ee
%
In flat gauge, we can identify the Sasaki-Mukhanov variable with
the Higgs field. The action for the Sasaki-Mukhanov variable is again that of
a canonically normalised and time-dependent harmonic oscillator
\eref{S_ho}, but now with effective mass:
\begin{align}
 m^2_v &=- \frac{z''}{z} = -H^2 \( 2+5 \eps_H - 3 \eta_H + (
6\eps_H^2-4\eps_H \eta_H -\xi_{(2)} )\) ,
\nn \\
&\to -H^2 \(2-3 \eta +5 \eps 
 + \mcO(\eps^2,\eta^2) \).
\label{m2}
\end{align}
 This is to be
compared with \eref{m1}, for the case of a probe field.  The
difference between including and neglecting gravity's back reaction is
of order $\eps$. For Higgs inflation $\eps \sim \delta^2$ and $\eta
\sim \delta$, with the expansion parameter $\delta$ defined in
\eref{delta}. To lowest order in $\delta$ we can therefore ignore the
back reaction.  As discussed in Sec.~\ref{sec:nondynamical}, this is
not the case if inflation is analysed in the Jordan frame.


\section{Calculation of the beta functions}
\label{sec:beta}

Our aim in this section is to compute the beta functions for the
couplings $\lambda$ and $\xi$.  We ignore the parameter $v$ in the
Higgs potential as it is negligibly small and plays little role at
energies above the electroweak scale.  The beta function, and hence
the RGEs, can be derived from the UV divergent part of the effective
potential, and is the approach we adopt here.  A previous set of
works~\cite{MP,GMP} shows how one can calculate the effective
potential by keeping the classical Higgs field $\phi_\text{cl}$
constant, and time-dependence, if necessary, can be incorporated by
simply setting $\phi_\text{cl} \to \phi_\text{cl}(t)$.  The result is
then the usual Coleman-Weinberg (CW) potential \cite{CW_paper}.  With
the effective potential and RGEs at hand one can then construct the
renormalisation group (RG) improved effective action (see
e.g. \cite{sher}).

We perform the calculation in the Einstein frame, for two main
reasons.  First, most of our intuition and standard results are valid
in this frame, where gravity is minimally coupled.  In the Einstein
frame there is a clear interpretation of mass scales, in contrast with
the Jordan frame where field-dependent measuring sticks are used (due
to a field-dependent Planck mass).  Second, in the Einstein frame
gravity can be treated as a classical background, as discussed later
in Sec.~\ref{sec:nondynamical}, and we do not need to worry about a
strong coupling between quantum degrees of freedom which correct the
classical potential, and gravity perturbations.

Because the potential in the Einstein frame is non-renormalisable we split
the running into the small, mid- and large field regimes and consider
each regime as a separate effective field theory (EFT).  As long as energies
stay below the cutoff of the EFT we can renormalise the couplings and
compute the beta functions in each regime, and patch them together to
obtain a final answer.

The CW corrections to the potential depend on the masses of the fields
running in the loop; in our case we have only the Higgs.  As computed
in Ref.~\cite{GMP}, treating gravity as a classical background, the
FLRW space-time corrects the masses in the loop by an amount
proportional to $\dot{H}+2H^2$.  In Fig.~\ref{F:mass} we plot the
Hubble scale $H^2$ and Higgs mass squared, under the assumption that
the Higgs field dominates the energy density (which is valid during inflation and gives only a negligible error afterwards).   As can be seen,
during inflation the Hubble scale dominates the Higgs mass and so we
need to take the FLRW quantum corrections into account.  After
inflation these can be neglected and a Minkowski calculation suffices.

Our aim in computing the beta functions is to relate low and high
scale observables by running the couplings in between the electroweak
and inflationary scale.  To that end we assume that the running is
path independent, that is, independent of the precise details of how
the inflaton (the Higgs in our case) rolled down the potential and the
universe was reheated.  As long as the time-evolution is (close to)
adiabatic it should be safe to make such an assumption.

We use dimensional regularisation with MS-bar renormalisation
prescription.  We do not include the anomalous dimension, as it
vanishes and so does not contribute to the running.

\subsection{Higgs mass in the Einstein frame}

For the loops, we need to compute the physical Higgs mass in the
Einstein frame.  With non-canonical kinetic terms, masses can be
computed using the covariant generalisation of
$m_\phi^2=\partial_\phi^2V$, namely
\be
(m^2)^i_j = \gamma^{ik} D_k D_j V(\phi) 
= \gamma^{ik}(\partial_k \partial_j V - \Gamma^l_{kj} \partial_l V) , \label{masscomp}
\ee
with $\gamma_{ij}$ the field space metric, and $\Gamma^i_{jk}$ the associated
connection coefficients. 
For now we take the Higgs as a real scalar (we treat the complex case
in Sec.~\ref{sec:complex}), which simplifies \eref{masscomp} and gives for the
Higgs mass (using $\xi \gg 1$)
\be
m_h^2 = 3 \lambda \phi^2
\frac{(1+4 \xi^2 \phi^2 -4 \xi^3\phi^4 )}
    {\Omega^4 (1+6\xi^2 \phi^2)^2}
\approx \{
  3 \lambda \phi^2,\; \frac{\lambda}{3\xi^2},\;
  -\frac{\lambda}{3 \xi^3 \phi^2}
\} ,
\label{mass}
\ee
where the second expressions are the leading terms in the three field
regimes.  The Higgs mass squared is negative during inflation. The
potential is convex, leading to a red tilted spectral index in
excellent agreement with the data. (Note however that these masses are computed in a Minkowski background. In the large field regime we will have to take the expansion of the universe into account, which leads to extra FLRW-corrections.)

\subsection{Small field regime} 

To obtain the dominant terms responsible for the running in
the small field regime we expand the action in the small parameter
$\delta \equiv \xi \phi \ll 1$.
At lowest order the potential reduces to the familiar $\phi^4$
theory.  Indeed, to lowest order the canonically renormalised field is
$h = \phi$, and the classical potential is $V^\text{cl} = \lambda \phi^4/4$.
The non-minimal coupling $\xi$ drops out completely. The CW correction
to the potential is \cite{CW_paper}
\be
\delta V = \frac1{64\pi^2} \sum_i (-1)^{f_i} m_i^4 \[ \ln\frac{m_i^2}{\mu^2}
-c_i \]
\label{CW}
\ee
with $f_i = 1 (-1)$ for a boson (fermion), and $c_i =3/2$ for a
fermion and scalar boson, and $c_i=5/6$ for a vector boson in the
$\overline{\rm MS}$ scheme; $\mu$ is the renormalisation scale. For
the Higgs field the effective mass in the small field regime is
$m^2_{h} =3\lambda \phi^2$, see \eref{mass}.  The beta functions
are found by demanding the full potential to be independent of the
arbitrary renormalisation scale, $\mu ({\rm d}V/{\rm d}\mu)=0$, giving
\be
\beta_\lambda \partial_\lambda V^{\rm cl} = -\mu\partial_\mu \delta V,
\ee
with $\beta_\lambda \equiv \mu \partial_\mu \lambda$.  This gives the
standard result
\be
\beta_\lambda  = \frac{9}{8\pi^2} \lambda^2.
\label{b_small}
\ee

The theory is renormalisable in the usual EFT sense: higher order
terms are irrelevant operators and can be neglected at low scales.
However, we can improve the calculation by including the next order
in $\delta$.  This introduces the leading irrelevant operator in the
tree-level potential and its coefficient includes the non-minimal
coupling $\xi$, allowing us to compute the running of this coupling.
To this next order we find $\phi = h -\xi^2 h^3 + \mcO(\delta^4)$ and
\be
V^\text{cl} = \frac{\lambda}{4} h^4 -\lambda \xi^2 h^6 + \mcO(\delta^4) .
\label{v_small}
\ee
The leading term that we have ignored in $V$ goes like $\delta^4 h^4$,
which is subdominant to the sextic term $\lambda\xi^2h^6 \sim \delta^2 h^4$.
Using \eref{v_small} we can calculate the effective Higgs mass and thus
the CW potential.  Neglecting $h^8$ terms in the 1-loop potential (higher
order counter terms need to be introduced to absorb this divergence)
we can derive the beta functions for $\lambda$ and $\xi$ from the quartic
and sextic operators via
\be [\beta_\lambda \partial_\lambda + \beta_\xi \partial_\xi] V^\text{cl}
    = - \mu \partial_\mu \delta V
\ee
The beta function for the coupling is as before \eref{b_small}; in
addition we find
\be
\beta_\xi = \frac{9}{4\pi^2} \lambda \xi .
\label{xi_small}
\ee
This result is only of theoretical importance, as the effect of
$\xi$, even though the coupling is large, is too small at electroweak scales
for it to be measurable.  Putting back powers of $\mpl$ we see
that the coefficient of the $h^6$ term in \eref{v_small} is
$-\lambda\xi^2/\mpl^2$, corresponding to a scale roughly $10^{15}$ GeV,
and is thus not measurable by colliders in the foreseeable future.
We cannot therefore probe the running of $\xi$ experimentally in the
small field regime.  On the other hand the coupling $\lambda$ can be
be independently measured, through the $h^4$ operator.  Our main point
here is that although to lowest order $\xi$ drops out, it still runs in
the small field regime.  Indeed, the running is not suppressed by large
scales and does not vanish in the small field regime.

\subsection{Large field regime}

We look next at the running in the large field regime, where the
expansion parameter is $\delta \equiv 1/(\xi \phi^2) \ll 1$.
In the slow roll approximation the effective mass of the adiabatic mode
which runs in the loop is, from \eref{m2}, 
$ m_h^2 =-H^2(2-3\eta +5 \eps)$.  

For technical reasons it is easier to work at the level of the equation
of motion, rather than of the effective action directly.  We use flat gauge,
where we can identify the Sasaki-Mukhanov variable with the Higgs field.  The
1-loop corrected equation of motion for the canonically normalised
field $h$ is (see Eq.~(35) of Ref.~\cite{sloth})
%
\be
0=\ddot h +3 H \dot h + V_h + \(\frac12 V_{hhh} + \frac34 \frac{\dot
  h}{H} H^2 (3\eta -2 \eps) \)G_h(0) ,
\label{eom_large}
\ee
with the equal-time propagator
\be
G_h(0) = \frac{1}{16\pi^2} m_h^2 \( \ln \frac{ m_h^2}{\mu^2}
-\frac32 \) ,
\label{G0}
\ee
where $\mu$ is the renormalisation scale.  As before we work with the
coupling $\bar \lambda \equiv \lambda/\xi^2$; the classical potential
is only a function of $\bar \lambda$, and thus so are the quantum
corrections; see \eref{large_Vh}.

To find the beta functions we require the effective potential to be
independent of the renormalisation scale.  Working at the level of the
equation of motion avoids having to integrate to get the action, and
independence of $\mu$ implies
\be
\beta_{\bar \lambda} \partial_{\bar \lambda} V_h^{\rm cl} =
-\( \frac12 V_{hhh} + \frac34 \frac{\dot
  h}{H} H^2 (3\eta -2 \eps) \) \partial_{\ln \mu}  G_h(0).
\label{eomh}
\ee
The last term in \eref{eom_large} and \eref{eomh} comes from
treating gravity dynamical, and is otherwise absent.  This correction is
subdominant in the expansion parameter $\delta$.  Moreover, the correction
to the mass (and thus the propagator) due to back reaction is order $\eps$,
which is also negligible to lowest order (since $\eta \sim \delta$ and
$\eps \sim \delta^2$).

To evaluate $\beta_{\bar \lambda}$ we eliminate the Hubble parameter
in \eref{eomh} using the background equations of motion $\dot
h^2=2\eps H^2$ and $H^2 = V(1+\eps/3)/3$, and thus $\dot h/H =
-\sqrt{2\eps}$ (the minus sign because the field is rolling down).  We
then use \erefdbl{large_Vh}{large_sr} to trade the potential
and its derivatives, and the slow roll parameters, in terms of an
expansion in $\delta$.  This gives the 1-loop corrections to the
equation of motion, and at lowest order in $\delta$ we find:
\be
\beta_{\bar \lambda} = -\frac{\bar \lambda^2}{144\pi^2} + \mcO(\delta).
\label{b_bar_large}
\ee

The minus sign comes
about because $ m_h^2$ is negative, while $V_{hhh}$ is positive
(usually, for minimally coupled fields, one gets a factor
$(\partial_{h_0}  m^2)  m^2 >0$, and the quartic coupling has
a positive-sign beta function).  Note in this respect that when Hubble
corrections dominate the mass term, the expression $\delta V \sim m^4
\ln \mu$ is no longer a good approximation.  The Hubble corrections to
the mass come from the kinetic terms (they are derivatives of the
scale factor) and only show up in the quadratic action.  The cubic
action remains unaltered in an expanding universe.  Hence, when
calculating the tadpole diagram, the result is as in
\eref{eom_large}.

To get the lowest order result in slow roll parameters (more precisely,
in $\delta$) one can treat gravity as a classical background, since,
as shown in Sec.~\ref{sec:backr}, in the Einstein frame dynamical gravity
introduces corrections of order $\delta^2$.  Higher order corrections should
only give consistency conditions.  If the theory is renormalisable\footnote{Note that in all of this section by ``renormalisable" we mean ``renormalisable in the effective field theory sense," i.e. that we can expand the Lagrangian in powers of $\delta$ and that the divergencies can be absorbed in the counterterms order by order.}, and so
only a finite number of counter terms are needed to absorb all divergencies,
the divergencies found at next order in $\delta$ should automatically be absorbed.
To check this, we also calculated
the beta function at next order, which includes the back reaction from
gravity.  We find indeed a consistent result, specifically, \eref{b_bar_large}
is correct to $\mcO(\delta^2)$.  As discussed in \cite{bezrukov:consistency}
the renormalisability of the theory can be understood in that the
potential in the large field regime asymptotes to a constant, and has
an approximate shift symmetry. This means that quantum corrections
should also respect the approximate shift symmetry, and thus be of the
same form.  This assures that they can be absorbed in the counter terms
of the classical potential.

To compare the large-field beta function with the beta functions calculated
in the small field regime, we find an expression for $\beta_{\bar\lambda}$
valid for small field:
\be
\beta^\text{small}_{\bar\lambda} = -\frac{27}{8\pi^2}\bar\lambda^2\xi^2 .
\ee
Note that the numerical coefficient in \eref{b_bar_large} is much smaller
than in the small field regime, and it is also down by the large factor $\xi^2$.
The running is thus relatively slow in the large field regime, which is as expected
since the potential asymptotes to a constant there.

\subsection{Mid-field regime}
\label{sec:midfield}

The issue of renormalisability becomes acute in the mid-field regime.
In the small field regime renormalisability was assured because higher
order operators are irrelevant in the IR limit, whereas in the large field
regime the shift symmetry came to the rescue.  In the mid field regime we
have neither, and as a consequence the best we can do (without specifying
any UV completion of the model) is to put a bound on bound on how big the
corrections will be.

In the mid field regime the expansion parameters are $x \equiv \xi
\phi^2 \ll 1$ and $y \equiv \xi^2 \phi^2 \gg 1$, and we expand in both
(taking them both equally small, i.e.\ write $x \to \delta x$ and $1/y
\to \delta/y$ and expand in $\delta$).  This is only a good approximation in
the middle of this regime, where both expansion parameters are order $0.1$.
Equivalently, one can take the expansion parameter to be $\delta=1/\sqrt{\xi}\ll1$,
so that $\xi=\delta^{-2}\tilde\xi$ and $\phi=\delta^{3/2}\tilde\phi$, where
$\tilde\xi$ and $\tilde\phi$ are order 1 quantities.  Then the boundaries
of the mid-field regime are $x=\delta\tilde\xi\tilde\phi^2$ and
$1/y=\delta/\tilde\xi^2\tilde\phi^2$, and as $\delta\to0$ these bounds
approach $(0,\infty)$.

Since the Hubble parameter drops fast below the Higgs mass in this
regime (see Fig.~\ref{F:mass}), we can neglect gravity corrections and do
a Minkowski calculation.

Expanding in $\delta$ the Einstein frame potential and the effective Higgs
mass~\eref{mass} yields
\begin{align}
V &= \frac{\lambda}{4} \delta^6 \phi^4
  \(1 - 2\delta\xi\phi^2 + 3\delta^2\xi^2\phi^4 - 4\delta^3\xi^3\phi^6\)
  + \mcO(\delta^{10}),\\
m_h^2 &= \frac{\lambda}{3\xi^2} \delta^4
  \(1 - \frac{\delta}{12\xi^2\phi^2} - 3\delta\xi\phi^2 \)
  + \mcO(\delta^6).
\end{align}
Using the CW-potential \eref{CW}, and demanding it to be independent
of the renormalisation scale,
\be
(\beta_\lambda \partial_\lambda + \beta_\xi \partial_\xi) V
= - \partial_{\ln \mu} \delta V,
\ee
gives\footnote{Note that $\beta_\xi$ can be negative order in $\delta$
so we must keep higher powers of $\delta$ for its contribution here.}
\begin{align}
&\frac14 {\beta_{\lambda} \delta^6\phi^4}  \(1-\delta 2 \xi \phi^2 + \delta^2 3 \xi^2 \phi^4-
\delta^3 4 \xi^3 \phi^6\)- 
\frac12{\beta_\xi \delta^9\lambda \phi^6} \(1-3\delta \xi \phi^2 +...\)
\nn \\
&\hspace{1cm} =\frac{\delta^8}{32\pi^2} \(\frac{\lambda}{3\xi^2}\)^2 
\(1-\delta \( \frac{1}{6\xi^2\phi^2} +6\xi \phi^2\) \).
\label{betas_mid}
\end{align}
The left- and right-hand side have different field dependence.  This
indicates that the theory is non-renormalisable as the 1-loop
corrections (right) cannot be absorbed in the counter terms of the same
form as the classical potential (left).  We need extra UV
physics/counter terms to absorb the divergencies coming from the
quantum corrections.  The crucial point, however, is that this UV
physics only needs to come in at second order in $\delta^2$.  This
allows us to hope that at lowest order the theory still works as an
effective theory, with any unknown UV corrections only appearing at
higher order.  Working under this assumption allows to find the lowest
order beta functions.

The leading order term, of order $\delta^8$, in the right-hand side of
\eref{betas_mid} induces a constant.  If we are conservative
and regard this constant term as requiring a new counter term as well
then the best we can do for the beta functions is
\be
\beta_\lambda = 0 + \mcO(\delta^2), \quad
\beta_\xi = \mcO(\delta^{-1}) .
\ee
If, on the other hand, the constant term can remain an induced cosmological
constant, we obtain
\be
\beta_\lambda = 0 + \mcO(\delta^3), \quad
\beta_\xi = \mcO(\delta^0) .
\ee
Whilst the expressions only give an answer that is zero to the order indicated,
they are nonetheless useful, as they guarantee that corrections to the 
running of $\lambda$ and $\xi$ from unknown UV physics will only enter at such
an order.

To compare with the small and large field regime we compute the beta
function for $\bar\lambda$.  For the conservative case we have that
$\beta_{\bar\lambda}=\mcO(\delta^5)$, and for the case where we allow an
induced cosmological constant, $\beta_{\bar\lambda}=\mcO(\delta^6)$.
If we write these in terms of $\xi=\delta^{-2}$ and explicitly put
in a factor $\bar\lambda^2$ to compare with the other regimes, we
have $\beta_{\bar\lambda}=\bar\lambda^2\mcO(\xi^{3/2})$ and
$\beta_{\bar\lambda}=\bar\lambda^2\mcO(\xi)$ respectively for the two cases.
In terms of magnitude of the beta function for $\bar\lambda$, this
result in the mid-field regime interpolates nicely between the small
and large field regimes.

\subsection{Summary}

We have calculated the beta functions for $\lambda$ and $\xi$.  In the
small field regime we find expressions for both, although the running
of $\xi$ here is only of theoretical importance.  The mid-field regime
is non-renormalisable and the best we can do is provide an upper bound
for the magnitude of the beta functions in terms of the controlling
expansion parameter $\delta$.  Although it is not possible to be more
precise in this regime without specifying the UV completion of the
theory, the running has a weak dependence on the unknown UV
operators.\footnote{The Higgs (and GB, see Sec.~\ref{sec:complex})
  contributions to the mid-field regime are negligible in comparison
  to additional gauge fields and fermions.  The dependence on the UV
  completion is therefore even less important in the more realistic
  case of the standard model, a point we intend to explore in a
  subsequent paper.}  In the large field regime we obtain the running
only for the combination $\bar\lambda \equiv \lambda/\xi^2$, and so we
are free to choose (in this regime only) the running of an independent
combination of $\lambda$ and $\xi$.  Although we calculated our result
in the Einstein frame, the same beta functions are valid also in the
Jordan frame, as we shall discuss in detail in the following section.

Without loss of generality, we can take $\xi$ a constant in the large
field regime (that is, the beta function vanishes there) and put all the
running in $\lambda$.  Then, for the large field,
$\beta_{\bar \lambda} = \beta_\lambda/\xi^2$ and we can give the
beta function over the whole range:
\be
\beta_\lambda = \lambda^2 \times
\left \{\frac{9}{8\pi^2},\; \mcO(\xi^{-1}),\; -\frac{1}{144\pi^2\xi^2}\right \} ,
\label{betalam}
\ee
for the small, mid-, and large field regimes, taking the conservative
results for the mid-field.  For completeness, we also state the
corresponding full beta function for $\xi$:
\be
\beta_\xi = \left \{\frac{9}{4\pi^2}\lambda\xi,\; \mcO(\xi^{1/2}),\; 0 \right \} .
\ee


\section{Compatibility of Jordan and Einstein frames}
\label{sec:jvse}

Both the Jordan and Einstein frame appear naturally in the analysis of
Higgs inflation. This raises the question whether there is any
difference between the frames, whether for example the final results
for the RGEs depend on frame-specific choices.  The debate in the
literature has not fully settled yet, but there seems to be consensus
that the answer is yes. Two different aspects come to the fore.
First, how does one regularise and normalise the loop corrections in Higgs
inflation?  More specifically, should one use a constant cutoff and
renormalisation scale in the Jordan frame or rather in the Einstein
frame?  Second, how is the theory quantised: can gravity be treated as
classical background or must it be included in the quantisation?  In
this section we will discuss these two aspects in turn.  Our claim is
that the two frames are fully equivalent and lead to the same physical
results when the calculation is done carefully.

The equivalence of the Jordan and Einstein frame is to be expected. At
the classical level the conformal transformation of the metric
\eref{Omega} just comes down to a field redefinition. Note that this
redefinition is not a symmetry transformation unless the action enjoys
a conformal symmetry, which it does not in general, and not in Higgs
inflation.  Just as physical results are the same whether they are
calculated using Cartesian or polar coordinates, it also does not
matter whether Jordan or Einstein frame fields are used. At the
classical level it can be shown explicitly that the two frames are
related by a 1-to-1 mapping of the fields, and thus lead to the same
physics. At the quantum level they give the same results for the 2-
and 3-point function~\cite{janW1,janW2,white1,white2}\footnote{We discuss the
  subtleties regarding quantisation in subsection
  \ref{sec:nondynamical}.}.

A more intuitive understanding of why the two frames are equivalent is
that all the conformal transformation does is scale all length scales,
or equivalently, all mass scales in the system. No physical experiment
is sensitive to this overall scaling.  Instead, what is useful, well-defined
and measurable are mass ratios (and length ratios, etc.) such as the proton
to the Planck mass or the proton to electron mass.
These remain invariant under a conformal rescaling of the metric.

Let us elaborate a bit more on the scaling of length and mass
scales. The transformation to the Einstein frame keeps the coordinates
the same (i.e.\ two events have the same coordinates in both frames)
but redefines the metric, and hence the line element used to measure
distances.  In the Jordan frame the metric is $g^J_{\mu\nu}$ and so
distances are measured by $ds_J^2 = g^J_{\mu\nu} dx^\mu dx^\nu$.  In
the Einstein frame the metric is $g^E_{\mu\nu}=\Omega^2g^J_{\mu\nu}$
and the line element is $ds_E^2 = g^E_{\mu\nu} dx^\mu dx^\nu =
\Omega^2ds_J^2$.  The line elements are therefore different, but this
is not surprising because a conformal transformation is a local change
of scale.

The crucial point is that the metric allows one to measure distances
which are relative distances, relative compared to the Planck scale.
Defining the effective Planck mass from the coefficient of the Ricci
scalar in the action, $\mcL \supset \sqrt{-g} \frac12 \mpl^2 R$, gives $\mplvar{J}^2 = \mplvar{E}^2 \Omega^2$.   Since the line element is dimensionful, a
change of the Planck scale (which occurs when going to the Einstein
frame) implies a change in the units of the line element as well.

Taking the last two paragraphs together shows that the invariant quantity
under the frame transformation is the physical distance in Planck units,
namely
\be
\mplvar{J}^2 ds^2_J = \mplvar{E}^2 ds_E^2
\ee
Therefore, the physical quantity is the dimensionless $\mpl^2 ds^2$ which is,
as all dimensionless ratios, equivalent in the two frames.

As length is scaled by the conformal transformation so are all mass
scales:
\begin{align}
\text{Jordan frame} &\quad\longrightarrow\quad \text{Einstein frame} \nonumber\\
m_J(\phi) &\quad\longrightarrow\quad \frac{m_J(\phi)}{\Omega(\phi)} = m_E(\phi)
\label{mass_transf}
\end{align}
We will show this scaling explicitly for a bosonic field in the next
subsection. As mentioned above, all mass ratios, such as $m_{\rm
  proton}/\mpl$ or $m_{\rm proton}/m_{\rm electron}$, are invariant
under the scaling.  In dealing with quantum fluctuations it is
important to realise that \emph{all} dimensionful quantities scale as
above, including the cutoff and renormalisation scale.  Ratios
constructed from mass scales from different frames, e.g.\ $m_{{\rm
proton},J}/\mplvar{E}$, are frame dependent; they are unphysical in the sense
that they do not correspond to measurable quantities.

Thus the Jordan and Einstein frames, or for that matter any other
frame related by a conformal transformation, are not special in any
way, at least from a physical point of view.  The former is where the
theory is defined, and the potential is a simple polynomial. However,
actual calculations in the Jordan frame are complicated by the non-minimal
gravity sector, and one must properly take into account the mixing between
the Higgs and the metric degrees of freedom.
The Einstein frame is special in that the gravity sector is minimal. But
here too there is a price to pay, as calculations are now complicated by
the non-minimal kinetic terms of the Higgs.  Which frame to use for
calculations is fully optional, and the physical results should not depend
on the choice.

In the next subsection we will compute the effective potential in both
frames and show its equivalence, at least to 1-loop.  This result is
in contrast to the myriad of existing results in the literature, for
example those found
in~\cite{bezrukov_1loop,tranberg,buchbinder,kirsten,barvinsky,barvinsky3,lerner}.
We do the computation using cutoff regularisation, with a constant
cutoff in either frame, and show that the dependence on the choice of
the cutoff does not change the final result.  The calculation is also
performed using dimensional regularisation, and we find that both
frames again yield the same effective potential. In subsection
\ref{sec:nondynamical} we discuss quantisation and the back reaction
of gravity.

\subsection{Cutoff regularisation and a field dependent cutoff} \label{cutoffreg}

When using cutoff regularisation naively there seems to be a choice
whether to use a field independent cutoff in the Jordan or
Einstein frame.  Either the cutoff is a constant $\Lambda$ in the
Jordan frame which transforms, by \eref{mass_transf}, to the
field-dependent quantity $\Lambda/\Omega(\phi)$ in the Einstein frame.
Or the cutoff is $\Lambda$ in the Einstein frame, transforming to
$\Lambda\Omega(\phi)$ in the Jordan frame.  Which prescription to take
is a point of debate in the literature as the two choices seem to
yield different results~\cite{bezrukov_1loop,bezrukov_2loop}. 

Taken at face value, the above statements go against the
well-established concept of decoupling in QFT, which states that
physics at different energy scales decouple.  In particular, one can do
a low scale calculation without knowing the UV physics, and in this
case without knowing at what scale new degrees of freedom become
important (if we define this as the cutoff scale).  If decoupling
breaks down for Higgs inflation then it is impossible to make unique
predictions without knowing the UV physics.  In the following we
shall show that one can take a constant cutoff in either frame and
still obtain the same result, demonstrating that decoupling does not
break down.

Consider doing the 1-loop calculation in the Jordan frame, using a
constant cutoff in this frame.  Assume for the moment that the results
transformed to the Einstein frame indeed give rise to a field dependent
cutoff.  However, this does not mean the usual Einstein calculation
with a constant cutoff is incompatible with this Jordan-frame based
result.  Indeed, we can always choose another, constant cutoff which
lies below this field dependent cutoff.  Then the usual calculation
goes through, where we can formally send the cutoff to infinity, and
all dependence on it drops out.  In particular, it implies low scale
physics is not sensitive to the high scale field dependent cutoff that
is a remnant of computing the loops in the Jordan frame.  This is at
odds with the statement that the calculation with a constant cutoff in
the Jordan or Einstein frame gives different result.

The resolution is that both frames with a constant cutoff give the
same physical results. The point is that the effective potential does
not depend on the cutoff but rather on the cutoff divided by a mass
scale. That is, it depends on a dimensionless ratio which is invariant
under a conformal transformation.  We will show this explicitly.

We ignore the issue of canonically normalising and quantising the
gravity-Higgs sector, in that we do not consider Higgs quanta running
in the loops for the CW corrections to the potential.
We consider only external particles running in the loops, and since
they do not (by assumption) mix with the metric degrees of freedom we
can safely quantise them in either frame and keep the gravity-Higgs
sector a non-dynamical, constant background.  This simplification does
not change the essence of the problem as there remains the issue of
transforming the effective potential from one frame to the other.  We
will use a cutoff regularisation as it has a more direct physical
interpretation of the scales involved and clarifies the prescription
for transforming from one frame to the other. It also allows for an easy
comparison with the literature.  At the end we will comment on how
the calculation proceeds using dimensional regularisation.

Consider the Jordan frame Lagrangian for the Higgs radial component $\phi$
and an additional scalar $\chi$:
\be
\frac{\mcL_J}{\sqrt{-g_J}} =
\frac12 (1+\xi \phi^2)R + \frac12 (\partial\phi)^2 +\frac12 (\partial \chi)^2  -
V(\phi)  -\frac12 y^2 \phi^2 \chi^2 ,
\label{e:jordanlag}
\ee
where the Planck scale has been set to unity and $y$ is a coupling
constant.  After a conformal transformation $g^E_{\mu\nu}=\Omega^2g^J_{\mu\nu}$
with $\Omega^2=(1+\xi\phi^2)$, the corresponding Einstein frame Lagrangian is
\be
\frac{\mcL_E}{\sqrt{-g_E}} =
\frac12 R + \frac12\gamma (\partial\phi)^2 +\frac12 \frac{(\partial \chi)^2}{\Omega^2} -
\frac{V(\phi)}{\Omega^4}  -\frac12 y^2 \frac{\phi^2}{\Omega^2} \frac{\chi^2}{\Omega^2} .
\ee
$\gamma$ is a function of $\phi$ and defines the field-space metric; see \eref{non_can}.
Focussing on the extra scalar, we can introduce a canonically
normalised field $\tau = \chi/\Omega$, and the action becomes
\be
\frac{\mcL_E}{\sqrt{-g_E}} =
\frac12 R  + \frac12\gamma (\partial\phi)^2 +\frac12 (\partial \tau)^2
  -\frac{V(\phi)}{\Omega^4} 
-\frac12 y^2 \frac{\phi^2}{\Omega^2} \tau^2 + \ldots .
\label{e:einsteinlag}
\ee
The terms we have left out are the difference between
$[(\partial\chi)/\Omega]^2$ and $[\partial(\chi/\Omega)]^2$,
which include 3- and 4-point interactions
involving at least one $\tau$ and are not important for our discussion.
Since we are anyway treating the gravity-Higgs sector a constant background,
\eref{e:einsteinlag} gives correctly the leading quadratic terms when $\phi$
(hence $\Omega$) is a constant.
Looking at the Yukawa term, we see explicitly the relation \eref{mass_transf}
relating the Jordan frame mass scale $m_\chi = y \phi$ with the Einstein frame
mass $m_\tau$ via $m_\tau = m_\chi/\Omega$.

Consider the 1-loop correction to the potential of the $\tau$-particle.
As mentioned previously, we will neglect the subdominant 1-loop contribution
of the Higgs field itself.  In the Einstein frame the effective potential is
\be
V_E = \frac{\lambda}{4\Omega^4} \phi^4
    + \frac{\delta \lambda}{4\Omega^4} \phi^4
    - c m_\tau^4 \ln\( \frac{\Lambda_E^2}{m_\tau^2}\) ,
\ee
with $\delta\lambda$ the counter term, $c$ some number, and $\Lambda_E$
the cutoff on Einstein frame
(Euclidean) momentum.  The mass scale $m_\tau$ enters the log as the lower
boundary of the momentum integral.  Equally, one could have done the
calculation in the Jordan frame, from the Jordan
Lagrangian~\eref{e:jordanlag}, with result
\be
V_J = \frac{\lambda}{4} \phi^4
    + \frac{\delta \lambda}{4} \phi^4
    - c m_\chi^4 \ln\( \frac{\Lambda_J^2}{m_\chi^2}\) .
\ee
Now $\Lambda_J$ is a constant cutoff on Jordan frame momentum.
Performing a conformal transformation on the Jordan-frame result to go
to the Einstein frame, using the scaling relation \eref{mass_transf},
the potential scales as $V \to V/\Omega^4$.
It is crucial to realise that the cutoff transforms in the same way as all other
dimensionful parameters in the theory, and so the argument of the log
is a mass ratio that is invariant when going from one frame to the other.
We see therefore that the results are equivalent, that $V_J$ transforms
precisely to $V_E$.  Starting with a constant cutoff in the Jordan frame,
the only way to end up with a seemingly field dependent cutoff in the Einstein
frame is to take a mass ratio of scales defined in different frames, such as
$\Lambda_J/m_\tau$; but such a ratio is frame-dependent and
unphysical.\footnote{The argument is the same as the statement that
  comoving momentum is bounded by a constant comoving cutoff $p_{\rm
    com}< \Lambda_{\rm com}$, and is equivalent to the physical momentum
  being bounded by a constant physical cutoff $p_{\rm phys} <
  \Lambda_{\rm phys}$.  Here, comoving and physical scales are related
  by the scale factor: $p_{\rm phys} = a(t) p_{\rm com}$, and similarly
  for the cutoff.  Working instead with mixed quantities $p_{\rm phys}
  < \Lambda_{ \rm com}$ is not useful, and obscures the true physics.}
  
  In other words, even if $\Lambda_J$ and $\Lambda_E$ cannot both be field independent, as they are related by the field dependent conformal factor $\Omega$, we have argued here that what matters in any computation is the ratio of the cut-off and a mass scale. Both the cut-off and this mass scale transform in the same field dependent way. As a result their ratio, which is the physical dimensionless quantity that we are after, is frame independent and can therefore be taken constant in both frames.

Instead of transforming the unrenormalised Jordan frame result, as we did
above, one could also first renormalise and only then transform. We will now show
that this still leads to the same results for both frames. To do so we
start off with a short recapitulation of renormalisation in the Einstein frame,
and will then compare to the results in the Jordan frame \cite{sher}.

\paragraph{Einstein frame.}
Use the renormalisation prescription\footnote{Note that we do not
  define $\lambda$ in terms of fourth derivative of potential, as is
  usually done.  Our prescription is such that when acted on the
  classical potential the coupling is extracted (the first term
  in \eref{lambdaM}). We further normalise at $\mu = m_\tau$ rather
  than the more common $\mu = \phi$ because in the large field regime
  $m_\tau \propto H \to $ constant, while $\phi \to \infty$ runs
  off; thus the mass gives a better definition of the energy scales
  involved.}
\be
\lambda(\mu) = \frac{4 \Omega^4}{\phi^4} V_E \bigg|_{m_\tau^2=\mu_E^2}
= \lambda + \delta \lambda -4 c y^4 \ln \(\frac{\Lambda_E^2
  }{m_\tau^2}\) \bigg|_{m_\tau^2=\mu_E^2} .
\label{lambdaM}
\ee
This defines the counter term $\delta\lambda$. For a constant Einstein
frame cutoff, $\delta\lambda$ is field independent, as it should be for a
renormalisable potential.  Putting $\delta\lambda$ back into the potential
gives
\be
V_E = \frac{\phi^4}{4\Omega^4}\[ \lambda -4 c y^4 \ln\( \frac{\mu_E^2
  }{m_\tau^2} \)\] .
\label{e:einsteinv2}
\ee
The log vanishes for $\mu_E = m_\tau$, the typical scale during inflation.
Since the (not calculated) higher order corrections scale with the same
log dependence, this choice for $\mu_E$ minimises these corrections and
thus minimises the error in the 1-loop approximation.
The beta function is found by either requiring the potential be independent
of the renormalisation scale, $\mu \partial V_E/\partial \mu =0$, or by
differentiating \eref{lambdaM} with $\beta_\lambda =\mu \partial
\lambda(\mu)/\partial \mu$.  The resulting RGE is
\be
\beta_\lambda = 8 c y^4, \qquad  \lambda(m_{\rm EW}) = \lambda_0 ,
\label{beta_ex}
\ee
where we set the boundary condition at the electroweak (EW) scale.
Integrating the RGE, we can run the coupling from its known
value at the EW scale $\lambda_0$ to the inflationary scale (and
similarly for the Yukawa coupling $y$), i.e.\ we
integrate/run over the interval
\be
m_{\rm EW} < \mu_E < m_\tau .
\label{e:einsteininterval}
\ee
The resulting coupling $\lambda(m_\tau)$ at the inflationary scale
can be used to get the 1-loop RG-improved potential
\be
V_E(m_\tau) = \frac{\phi^4}{4\Omega^4}\lambda(m_\tau) .
\label{e:einsteineff}
\ee
Note that the cutoff dependence has dropped out of
\erefdbl{e:einsteinv2}{e:einsteineff} and we can formally send it
to infinity.

\paragraph{Jordan frame.}
We repeat the above steps but in the Jordan frame, and at the end transform
the RG-improved potential to the Einstein frame.  The renormalisation prescription
is
\be
\lambda(\mu_J) = \frac{4}{\phi^4} V_J \bigg|_{m_\chi^2=\mu_J^2}
= \lambda + \delta \lambda -4 c y^4 \ln \(\frac{\Lambda_J^2
  }{m_\chi^2}\) \bigg|_{m_\chi^2=\mu_J^2} .
\label{lambdaM2}
\ee
For a constant Jordan-frame cutoff the counter term is field
independent, as it should be.  Using the above to replace
$\delta \lambda$ in the potential gives
\be
V_J = \frac{\phi^4}{4}\[ \lambda -4 c y^4 \ln\( \frac{\mu_J^2
  }{m_\chi^2} \)\] .
\label{e:jordanv2}
\ee
For $\mu_J = m_\chi$ the typical scale during inflation, the log vanishes,
and again the error from the higher-order corrections is minimised. The
beta function only depends on the coefficient in front of the log in
\eref{e:jordanv2}, and is the same as the Einstein frame calculation,
\eref{beta_ex}, but with a different boundary condition
\be
\beta_\lambda = 8 c y^4,
\qquad \lambda(m_{\rm EW}^J) = \lambda_0 .
\ee
Note that the physical EW scale is transformed via (the reverse of)
\eref{mass_transf} to its Jordan frame value (since $\Omega \approx 1$
in the small field regime $m_{\rm EW}^J \approx m_{\rm EW}$). We can run
the coupling using the RGE above from its known value at the EW scale
$\lambda_0$ to the inflationary scale:
\be
m_{\rm EW}^J < \mu_J < m_\chi
\label{e:jordaninterval}
\ee
The resulting factor $\lambda(m_\chi)$ at the inflationary scale is
used in the 1-loop RG-improved potential
\be
V_J(m_\chi) = \frac{\phi^4}{4}\lambda(m_\chi) .
\label{e:jordaneff}
\ee
As in the Einstein-frame calculation, here in the Jordan frame the
cutoff dependence has dropped out of the equations.

We can compare the results here with those in the Einstein frame.  To
transform the renormalised potential \eref{e:jordanv2} we apply
\eref{mass_transf} and take $V_J\to V_J/\Omega^4$.  As the
argument of the log is a ratio of masses it remains unchanged, and the
result, $V_J/\Omega^4$, is equivalent to the Einstein-frame
renormalised potential, \eref{e:einsteinv2}.  The beta functions
contain no mass scales and are manifestly equivalent in both frames.
On the other hand, the $\mu_J$ running interval
\eref{e:jordaninterval} is dimensionful; its boundaries must be
transformed, as well as the the renormalisation scale itself,
$\mu_J\to\mu_J/\Omega=\mu_E$.  Doing this makes the interval
equivalent to the Einstein-frame interval \eref{e:einsteininterval}.
Finally, the RG-improved potential in the Jordan frame
\eref{e:jordaneff} is transformed to match the Einstein expression
\eref{e:einsteineff} by dividing through by a factor $\Omega^4$, as
well as transforming the argument of the running coupling,
$\lambda(m_\chi)\to\lambda(m_\chi/\Omega)=\lambda(m_\tau)$.  We see
then that all quantities associated with the effective potential are
equivalent in both the Jordan and Einstein frames, and would also be
equivalent in any other frame related by a conformal rescaling of the
metric.


The equivalence of the frames is independent of the regularisation scheme
used.  For example, using dimensional regularisation the effective potential
in the Einstein frame before renormalisation is
\be
V_E = \frac{\lambda}{4\Omega^4}\phi^4
    + \frac{\delta\lambda}{4\Omega^4}\phi^4
    - c m_\tau^4 \left[
        \frac{2}{\varepsilon} + \ln\(\frac{\mu_E^2}{m_\tau^2}\)
        + \text{const.} + \mcO(\varepsilon)
    \right] ,
\label{e:einsteindimreg}
\ee
where $c$ is the same constant as before, and $\varepsilon=4-d$ with
$d$ the number of space-time dimensions.  The Jordan-frame calculation
goes through in a similar way, and one obtains an expression for the
unrenormalised potential which, when transformed using
\eref{mass_transf}, matches the Einstein frame expression
\eref{e:einsteindimreg}.  One can then renormalise using minimal
subtraction (or a variation thereof) and the counter term
$\delta\lambda$ subtracts off the divergent piece $2/\varepsilon$
along with the unimportant constant.  The results in the two frames
are expressions equivalent to those obtained using cutoff
regularisation, \eref{e:einsteinv2} and~\eref{e:jordanv2} for the
Einstein and Jordan frames respectively.  From here the calculation of
the beta functions and the RG-improved potential follow as before and
the results are equivalent.

To summarise, there is a 1-to-1 mapping between the Jordan and
Einstein frame results, not only of the (RG-improved) effective
potential, but also of the RGE equations and the running interval. The
important point to realise is that {\it all} dimensionful quantities
scale as per \eref{mass_transf}, including the cutoff and
renormalisation scale.  There is therefore no ambiguity in choosing
the cutoff.  UV physics decouples, as it should.

\subsection{Quantisation, and gravity as a non-dynamical background}
\label{sec:nondynamical}

At the classical level the Jordan and Einstein frame are related by a
field transformation. One may ask whether this equivalence is retained
after quantisation. Although we expect it to be so --- the conformal
transformation is only a scale transformation --- we note that this
has not been shown explicitly.  In the previous
subsection we showed how CW corrections can be computed in either
frame to give the same result, but this is predicated on having
already quantised the degrees of freedom, namely those running in the
loops.  The quantisation of additional degrees of freedom is straightforward,
but subtleties may arise in the quantisation of the non-minimally coupled
Higgs field itself.  Ideally one would like to define the canonical
conjugate fields of the Jordan frame fields, and quantise them in the
usual way $[\phi,\Pi_\phi] =i$. The quantum corrections thus obtained
should then be shown to match the Einstein frame calculation.

Previous attempts to quantise in
the Jordan frame \cite{janW1, janW2, kaiser,white1,white2} (where the last two address the case of multifield inflation) do the following.
Starting with the Jordan frame action, one defines Sasaki-Mukhanov
variables for which the kinetic term of the adiabatic mode (a mixture
of the metric scalar and the Higgs) is canonical, and subsequently
quantises this canonical field.  However, by doing field redefinitions
to go to this preferred variable, one is implicitly changing frames
and one can no longer claim that quantisation is done in the Jordan
frame, nor is it done in the Einstein frame.  There is a unique frame
where the adiabatic mode is canonically normalised, and consequently a
unique Sasaki-Mukhanov variable. One can then choose to express this
variable in terms of Jordan frame fields ($g_{\mu\nu}^J, \phi$) or
Einstein frame fields ($g_{\mu\nu}^E,h$); see
Refs.~\cite{janW1,janW2}.  But quantising the Sasaki-Mukhanov variable
expressed in these two different ways is {\it not} the same as
quantising in the frame associated with the fields (Jordan or
Einstein) being used in the expression.  With the unique quantisation
prescription of the Sasaki-Mukhanov variable, and given that the
frames are equivalent at the classical level, it naturally follows
that all quantum computations such as the 2- and 3-point function or
the 1-loop calculation come out unique.  Using this unique variable essentially
fixes the frame.  It is therefore not possible to draw any definite conclusions
regarding frame equivalence, or lack thereof, regardless of the quantities
used to express the unique answer in.

A related issue is
whether it is a good approximation to take gravity classical, thus
ignoring the mixing between the metric and the Higgs degrees of
freedom, and only quantise the Higgs field. In the literature this
approach has been used for the Jordan frame calculation
\cite{bezrukov_1loop,tranberg,buchbinder,kirsten,barvinsky,barvinsky3,lerner}. Also
the RGE for the non-minimal coupling used in
\cite{
wilczek} has been derived using this approach.
In calculating the back reaction of gravity one quantises the
Sasaki-Mukhanov variable, which is the adiabatic mode.  In quantising
this variable we have quantised (part of) gravity; this is fine
because gravity is an excellent EFT valid at energies up to and
including the scale of inflation, which is well below the Planck
scale.  Making the approximation to treat gravity non-dynamical
amounts to freezing out the quantum fluctuations associated with
metric degrees of freedom.  Applied independently to the Einstein and
Jordan frames the approximation that one makes is different, as it
corresponds to freezing the quantum fluctuations of different
combinations of the scalar degrees of freedom from the metric and the
Higgs.  Performing the calculation in the Einstein frame, we have
shown in Sec.~\ref{sec:backr} that the back reaction is small, order
$\eps$, so the approximation does not introduce a large error.
Freezing the metric in the Jordan frame, the calculation breaks down
in the large field regime because one freezes too much of the physical
degree of freedom that is the inflaton.

Treating the Ricci scalar as a classical background field in the
Jordan frame renders the non-minimal coupling term $\xi|\Phi|^2R$
simply a mass term for the Higgs, and hence $\xi$ (or rather
$\xi-1/6$) runs as a mass.  For Higgs inflation, this means using the
small field RGEs over the whole field regime, in addition to this
running of $\xi$.  This disagrees with the Einstein frame results
derived in Sec.~\ref{sec:beta}.  We would like to put forth some
further arguments as to why it is not a good approximation to treat
gravity non-dynamical in the Jordan frame, and why the Einstein frame
results can, in contrast, be trusted.

If gravity is treated as a classical background then the
distinction between the Einstein and Jordan frames is moot. The only
way to distinguish the two frames is by seeing in which frame the
kinetic terms of the graviton are canonical, which is not possible
if one freezes out the gravity kinetic term $R$.  Hence, the calculation
in the Jordan frame with $R$ a background field does not properly
reflect what it means to be in the Jordan frame.

Treating $R$ as a classical field already gives incorrect results at
the classical level.  In the Einstein frame the non-minimal kinetic
terms of the Higgs are essential to find a small mass, $\eta \ll 1$,
suitable for inflation. Transforming back to the Jordan frame, the
scalar degree of freedom in the Jordan frame metric is a mixture of
the Einstein metric and Higgs field. Treating $R$ as classical in the
Jordan frame fails to identify the kinetic terms properly ($R$ is
really a kinetic term).  Instead, it seems that $R$ now contributes to
the mass of the Higgs.  Consequently, one does not find a small Higgs
mass, and thus no inflation.

In the large field regime the Einstein frame potential only depends on
the parameter combination $\bar \lambda = \lambda/\xi^2$.  Hence, so
do the quantum fluctuations, and we indeed find a beta function for
$\bar \lambda$, see \eref{b_bar_large}.  Since the frames are
equivalent at the classical level, also in the Jordan frame the
physics only can depend on $\bar \lambda$.  Hence, obtaining separate
beta functions for $\xi$ and $\lambda$, as is found in those
treatments of the Jordan frame, is an inconsistent result.

Although it is easiest to see that gravity has to be taken dynamically
in the large field regime in the Jordan frame calculation --- the
metric degree of freedom mixes strongly with the Higgs field during
inflation --- also in the small field regime this has to be the case.
Taking the small field limit in the Einstein frame gives $\phi=h$ and
the action becomes completely independent of $\xi$ (although the small
field limit is defined by $1/\xi$, we remain independent of $\xi$ for
small enough $\phi$).  In the Jordan frame, however, one does not
attain $\xi$-independence in the small field limit, as $\xi R \phi^2$
is a mass term and thus a relevant operator in the small field limit
(it is the same order as the kinetic term for $\phi$).

The arguments up to now have focused on the kinetic structure in
the Jordan and Einstein frames, and are relevant for any model with a
non-minimal coupling to gravity.  When applied to the specific model of
Higgs inflation, the Jordan frame calculation raises further questions.
With a quartic potential, the perturbative calculation works in the
Jordan frame as long as the couplings are small $\lambda^2 \ll 1$.
One never encounters the problem that the divergences cannot be reabsorbed
in a finite number of counter terms, something that we found to break down
at higher order in the mid-field regime when calculating in the Einstein
frame.  Extending the Higgs to a doublet, as is required for a full
standard-model calculation, in the Jordan frame there is never a problem
with non-minimal kinetic terms that cannot be diagonalised, in sharp
contrast to the situation in the Einstein frame where the field space
metric $\gamma_{ij}$ can not be made diagonal.

Although our above arguments are all heuristic, we find them
compelling reasons for trusting the Einstein-frame calculation over
the Jordan one.  Furthermore, one can also explicitly compare the RGEs
found in the Jordan frame with gravity classical, see
\cite{bezrukov_1loop,tranberg,buchbinder,kirsten,barvinsky,barvinsky3,lerner},
with our full calculation done in the Einstein frame. Transforming
frames using the scaling relations \eref{mass_transf} the results do
not agree, and the differences are generically not small.


\section{Extension to a complex scalar field}  
\label{sec:complex}

We can extend our analysis and computation of the beta functions
to the case where the Higgs is a complex scalar with 2 degrees of
freedom.  This is a necessary step if one wants to tackle the full
case of the standard model.  The main difference going from a real
to a complex scalar is the appearance of the Goldstone boson (GB),
the angular degree of freedom, and its contribution to the loop
corrections of the potential.  Because the Higgs is not in its
minimum during inflation this GB is actually a ``massive angular
degree of freedom'', but, because it will eventually become a true
GB after the Higgs has settled in its minimum, we shall continue
to refer to it as such.

\begin{figure}[t!]
\begin{center}
\includegraphics[width=0.7\textwidth]{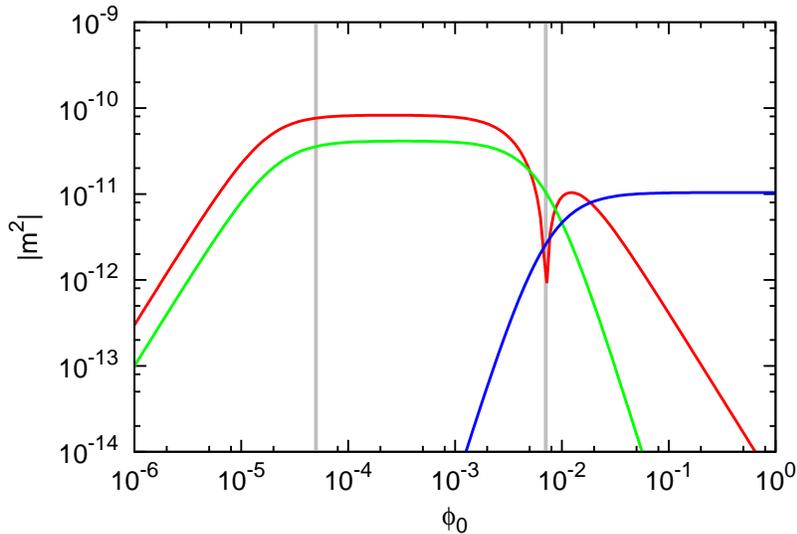}
\end{center}
\caption{Particle masses $|m^2(\phi_0)|$ as a function of the Higgs
vev $\phi_0$. Shown are the real Higgs mass (red) and the GB mass (green).
In addition the Hubble scale $H^2$ is indicated (blue).
The two vertical lines correspond to $\phi_0=1/\sqrt{\xi}$
and $\phi_0=1/\xi$ respectively.  Parameters used are $\lambda =0.1$
and $\xi =2 \times 10^4$.}
\label{F:mass}
\end{figure}

Due to the non-minimal kinetic terms, that cannot be rendered diagonal
by any field redefinition, we use \eref{masscomp} to find the mass
eigenvalues of the radial and GB components, which holds at any given
point in field space.  We find (again in the limit $\xi \gg 1$)
\begin{align}
m_h^2 &= 3 \lambda \phi^2 \frac{(1+4 \xi^2 \phi^2 -4 \xi^3\phi^4)}
{\Omega^4 (1+6\xi^2 \phi^2)^2}
\approx \{
3 \lambda \phi^2,\; \frac{\lambda}{3\xi^2} ,\; -\frac{\lambda}{3
  \xi^3 \phi^2} \} ,
\label{diagmass_h} \\
m_\theta^2 &= \lambda \phi^2 \frac{1}{\Omega^4
  (1+6 \xi ^2 \phi^2)} 
\hspace{1.15cm} \approx \{
\lambda \phi^2,\; \frac{\lambda}{6\xi^2},\; \frac{\lambda}{6\xi^4 \phi^4}\} ,
\label{diagmass_theta}
\end{align}
where the expressions on the right-hand side are the leading terms in
the three field regimes.  In the mid- and large field regimes we find
a $\xi$-suppression for the GB mass as well as the Higgs radial mode,
as opposed to what was found in
Refs.~\cite{barvinsky2,barvinsky3,lerner,allison} where the GB is not
suppressed.\footnote{Ref. \cite{barvinsky2} writes `Goldstone modes,
  in contrast to the Higgs particle, are not coupled to curvature, and
  they do not have a kinetic term mixing with gravitons.'  However,
  when the Higgs doublet is written using cartesian fields as per
  \eref{non_can}, it is clear all modes couple non-minimally.  Using a
  polar field decomposition instead, in the Einstein frame the kinetic
  terms for the GB look standard (after a rescaling of the field
  $\theta/\Omega \to \theta$) $\mcL \supset \frac12 \phi^2 \partial
  \theta^2$, but that is misleading, as $\phi$ itself is not the
  canonically normalised radial mode.}  This $\xi$-suppression can be
seen clearly in Fig.~\ref{F:mass}, where we plot the Higgs and GB
masses as a function of the background scalar value $\phi_0$.  We see
that $|m_\theta^2| < |m_h^2|$ almost everywhere, and in fact
$m_\theta^2\sim\phi^{-4}$ is highly suppressed compared to
$m_h^2\sim\phi^{-2}$ in the large field regime.  The GB thus gives a
subdominant contribution to $\delta V$ and it can be neglected to
first order.

For the RGEs, the results in the small and mid-field regime are easily
generalised to a complex scalar field.  The only thing to do is add the
contribution of the GB to the CW potential using the expression for the
mass, \eref{diagmass_theta}.  In the small field regime this gives
the replacements
\begin{align}
\beta_\lambda &= \frac{9}{8\pi^2} \lambda^2 \hspace{0.56em}\to \frac{9+1}{8\pi^2} \lambda^2,\\
\beta_\xi &= \frac{36}{16\pi^2} \lambda\xi \to \frac{36+3}{16\pi^2} \lambda\xi.
\end{align}
To obtain the expression for $\beta_\xi$ we again went to order $h^6$
in the potential.
For the mid-field regime, the CW corrections only enter at order
$\delta^2$ relative to the leading order term from the classical
potential (being conservative with the induced cosmological constant).
To lowest order we therefore find the same result as for the real scalar:
\be
\beta_\lambda = 0 + \mcO(\delta^2), \quad
\beta_\xi = \mcO(\delta^{-1}) .
\ee

In the large field regime we can work at the lowest order in the
expansion, neglecting the back reaction from gravity.  Both the
Higgs radial mode and the GB are light, with their mass dominated by
Hubble corrections and $\hat m_{h,\theta}^2 \approx -2 a^2 H^2$.
This implies that their propagators have the same form.  The equation that
determines the beta function, \eref{eomh}, generalises to
\be
\beta_{\bar \lambda} \partial_{\bar \lambda} V_h^{\rm cl} =
-\partial_{\ln \mu} \frac12 (V_{hhh} + V_{h\theta\theta})  G(0).
\ee
Since $V_{hhh} = \mcO(\delta)$ and $V_{h\theta\theta} = -2 \bar \lambda
\delta^2/(3\sqrt{6}) = \mcO(\delta^2)$, we can neglect the GB
contribution.  This could have been foreseen from Fig.~\ref{F:mass},
which shows that the GB mass is parametrically smaller in the large
field regime than the Higgs mass (and thus so is the derivative).  The
beta function in the large field regime is therefore the same for a real
and complex scalar.

In summary, the only modification to the beta functions due to the
inclusion of a GB is a change in $\beta_\lambda$ and $\beta_\xi$ in
the small field regime.  This change is easily generalisable to having
$n_\theta$ GBs (as needed for a Higgs doublet), and in this case the
running of $\lambda$ over the full range is
\be
\beta_\lambda = \lambda^2 \times
\left \{\frac{9+n_\theta}{8\pi^2},\; \mcO(\xi^{-1}),\; -\frac{1}{144\pi^2\xi^2}\right \} .
\ee

\section{On the unitarity bound}
\label{sec:unitarity}

There is a large and ongoing discussion in the literature regarding
the unitarity bound in Higgs inflation, see for example
Refs.
\cite{bezrukov:consistency,barbon,cliff1,cliff2,hertzberg,lerner1}.
The large non-minimal coupling $\xi$ leads to loss of unitarity of
$\text{SU}(2)$ gauge boson scattering, and this occurs at an energy
which depends on the value of the background Higgs field.  In the
small and mid-field regime the unitarity bound is below the Planck
scale, which suggests new physics, aside from quantum gravity, should
enter to restore unitarity at this scale. (Typical energy scales are however always below the scale of unitarity violation.)  The minimal assumption is a
strong phase of the theory \cite{bezrukov_2loop,bezrukov:consistency};
another option is adding new fields and couplings, see for example
Refs.~\cite{giudice,lerner2} for new interactions that restore
unitarity.  Whilst we do not have a complete answer to these issues,
we would like to make here a brief, and fairly speculative, comment
that the new physics necessarily appears only at Planck-scale energies,
and, as such, gravity may be enough to UV complete the theory.

Consider scattering two Higgs particles with high momenta while
keeping the background Higgs field in the minimum of the Mexican hat
potential, i.e., scattering at high energy while remaining in the
small field regime.  It is far from obvious you can do this.  Starting
in the small field regime, as you increase kinetic energy you are able
to, and in fact with quantum mechanical fluctuations must, probe the
large field regime of the potential.  This excites background field
quanta in the region of the hard scattering interaction point, and so
moves the theory into the large field regime at that point.

Following this reasoning, energy is equipartitioned over gradient and
potential energy, and it only makes sense to talk about scattering at
a given energy when the background field value is also of order this
energy --- rather than make a distinction between kinetic and
potential energy.  (In a QFT calculation it seems you can keep the
background $\phi$ fixed by hand; however, then this `equipartitioning'
should show up if you take higher order corrections into account.)
This means you can never probe the small field unitary bound, it is an
unphysical bound.  As you approach it you must take higher loop
corrections into account, and doing that properly it may turn out that
no new physics is needed to keep the theory unitary.  The real cutoff
of the theory can then be pushed to the Planck scale (since that is
the cutoff in the large field regime \cite{bezrukov:consistency})
where gravity corrections are assumed to UV complete the theory.


\section{Conclusions}
\label{sec:discussion}


In this paper we have performed a detailed analysis of the quantum
corrections and renormalisation in Higgs inflation, computing
the beta functions for the couplings $\lambda$ and $\xi$, and the
(RG improved) effective potential.  We have also given particular
attention to the differences between, and compatibility of, the
Jordan and Einstein frames.


Working in the Einstein frame, we have computed the beta functions of
$\lambda$ and $\xi$ in the small, mid- and large field regimes, finding results
that match consistently between the regimes; see Sec.~\ref{sec:beta}.
In the small field regime, $\lambda$ runs as it does in the standard model, and
we find the beta function for $\xi$ by examining the leading-order irrelevant
operator.  The mid-field regime is non-renormalisable and, whilst the beta functions
depend on the UV completion of the theory, we can still give conservative bounds
on them.  In the large field regime re\-nor\-ma\-li\-sa\-bi\-li\-ty is restored thanks to the
approximate shift symmetry and it is the combination $\lambda/\xi^2$ that runs.

The theory of Higgs inflation is not renormalisable in the Einstein frame,
and so by the assumed quantum-equivalence, the Jordan frame must also be
non-renormalisable.  In the Einstein frame one can keep gravity as a classical
background with minimal error, and treat the action as an EFT, so it is
in this frame that calculations are easiest.  Even though the theory is
non-renormalisable we are still able to make predictions for the beta
functions, as the higher order terms are negligible: in small field they
are suppressed in the IR; in mid-field they are not computable but are
bounded; in large field there is a shift symmetry.


In the literature an alternative way to compute the running uses the
s-suppression factor \cite{wilczek,lerner,allison}.  For this, one should
go to the Einstein frame, define $\pi = \partial L/\partial \dot \phi$, and
apply the usual commutation relations.  Note that the Einstein field $\phi$ here
does not have a canonical kinetic term.  One finds that
$[\phi, \dot \phi] = i s \delta(\vec x - \vec y)$.  This leads to the heuristic
that all Higgs propagators are suppressed by the field-dependent $s$-factor
\be
s(\phi) = \frac{1+ \xi \phi^2}{1+(1+6\xi) \xi \phi^2}.
\ee
In the small field regime $s\sim1$ and the running is as per the SM.
For large field, $s\sim1/6\xi$ acts as to suppress Higgs loops.
The CW potential is the calculated in the Jordan frame, using the
$s$-factor to account for the non-minimal kinetic terms of the Higgs.
While this prescription gives the correct qualitative behaviour~--- that
the contribution of the Higgs field to the effective potential is
suppressed in the mid- and large field regime~--- the exact expression
differs from our results.  Moreover, it is unclear how to incorporate
the GB in this prescription.

Our approach to including GBs in the effective potential and beta
functions is to compute their masses using the generalisation of the
second derivative of the potential, but in a curved field-space, \eref{masscomp}.
We have shown explicitly how this works by studying the extension of the
model to a complex scalar.
The GBs are massive because the Higgs is not in its minimum, and they
are suppressed during inflation, as is the Higgs radial mode.
In contrast, Refs.~\cite{barvinsky3,allison,lerner} claim that only the
Higgs field is be suppressed, not the GBs.


At the classical level, it is widely agreed that the Jordan and
Einstein frame are equivalent, as going back and forth between them
amounts to a redefinition of the fields (the metric and Higgs).
However, we also claim that when quantum corrections to the potential
are included both frames still give equivalent results for all physical
quantities.  This is in contrast to claims made in the literature.
In some of these works, Refs.~\cite{bezrukov_1loop,allison},
it seems that the disagreement follows from neglecting the transformation
of the renormalisation point $\mu$ and cutoff $\Lambda$.
In Refs.~\cite{bezrukov_1loop,barvinsky,barvinsky3,lerner,tranberg,buchbinder,kirsten}
it is the exclusion of gravitational back reaction in the Jordan frame
that we do not agree with.  Also in Ref.~\cite{
wilczek}
the RGE for $\xi$ is computed with gravity classical in the Jordan frame.
We claim that gravitational back reaction, computed in Sec.~\ref{sec:backr},
can only be safely ignored in the Einstein frame.


We note that a definitive proof of the quantum equivalence between the
Jordan and Einstein frames (or any other frame connected by a conformal
rescaling of the metric) should include a proper quantisation, of position
and conjugate momentum, in the Jordan frame, from which loop corrections
and $n$-point functions should be computed.  We are not aware of any work
that has carried out this task.  So far it seems that the quantisation has
always been of one-and-the-same Sasaki-Mukhanov variable.  However, we do not see
any reason to expect that such a proper quantum computation would reveal a
breakdown of the equivalence between the Jordan and Einstein frames.
Although our toy model in subsection \ref{cutoffreg} does not take Higgs
fluctuations into account, it demonstrates how to correct previous arguments
that make claims against quantum frame equivalence.


In regards to unitarity of Higgs inflation, we commented on the possibility
that, due to equipartitioning of energy over gradient and potential energy,
the new physics required to restore unitarity appears only at Planck-scale
energies.  As such, quantum gravity may be enough to keep the theory unitary
at all energies.


In this paper we have focused on a real Higgs field, with a brief
but important extension to the complex case to include GBs.  This
reduced model retained all the important physics and allowed us to
make our points above.
Building on this work, we now plan to address the case of the
renormalisation of the full standard model Higgs inflation in a
follow-up paper.


\section*{Acknowledgements}

DG is funded by a Herchel Smith fellowship.
SM and MP are supported by the Netherlands Foundation for
Fundamental Research of Matter (FOM) and the Netherlands
Organisation for Scientific Research (NWO). SM is also supported by the Conicyt ``Anillo" project (ACT1122).
DG would like to thank the members of the SUSY Working Group at
the Cavendish for useful discussions, and we thank our
cosmoleague Jan Weenink for numerous illuminating discussions.


\end{document}